\documentclass{IEEEcsmag}

\usepackage[colorlinks,urlcolor=blue,linkcolor=blue,citecolor=blue]{hyperref}

\usepackage{graphicx}
\usepackage{adjustbox}
\usepackage{booktabs}
\usepackage{pifont}
\usepackage{amssymb}
\usepackage{verbatim}
\usepackage{bibunits}
\usepackage{tcolorbox}
\usepackage{hyperref}

\usepackage{caption}
\usepackage{subcaption}

\usepackage{color}
\usepackage{amssymb}
\usepackage{grffile}
\usepackage{relsize}
\usepackage{colortbl}
\usepackage{enumerate}
\usepackage{amsmath}
\usepackage{tikz}
\usepackage{pgf}
\usepackage{pbox}
\usepackage{rotating}
\usepackage{multirow}
\usepackage{balance}
\usepackage[english]{babel}
\usepackage{tablefootnote}
\usepackage{flushend}

\usepackage[nocompress]{cite}

\usepackage{soul}
\usepackage{upmath}

\jvol{XX}
\jnum{XX}
\paper{8}
\jmonth{May/June}
\jname{IEEE Computer}
\pubyear{2019}

\begin{document}

\sptitle{Department: Head}
\editor{Editor: Name, xxxx@email}

\title{Explainable AI for Software Engineering}

\author{Chakkrit Tantithamthavorn}
\affil{Monash University, Australia.}

\author{Jirayus Jiarpakdee}
\affil{Monash University, Australia.}

\author{John Grundy}
\affil{Monash University, Australia.}

\markboth{Department Head}{Paper title}

\begin{abstract}
Artificial Intelligence/Machine Learning techniques have been widely used in software engineering to improve developer productivity, the quality of software systems, and decision-making.
However, such AI/ML models for software engineering are still impractical, not explainable, and not actionable.
These concerns often hinder the adoption of AI/ML models in software engineering practices.
In this article, we first highlight the need for explainable AI in software engineering.
Then, we summarize three successful case studies on how explainable AI techniques can be used to address the aforementioned challenges by making software defect prediction models more practical, explainable, and actionable.
\end{abstract}

\maketitle

\chapterinitial{AI for Software Engineering}

The success of software engineering projects largely depends on much complex decision-making~\cite{hassan2008road} (e.g., Which tasks should a developer do first? Who should perform this task? Is a software system of high quality? Is a software system reliable and resilient enough to deploy?). 
However, the erroneous of these complex decisions is costly in terms of monetary and reputation.

\begin{figure*}[t]
  \centering
  \includegraphics[clip,width=.9\linewidth, trim={0 0 0 0}]{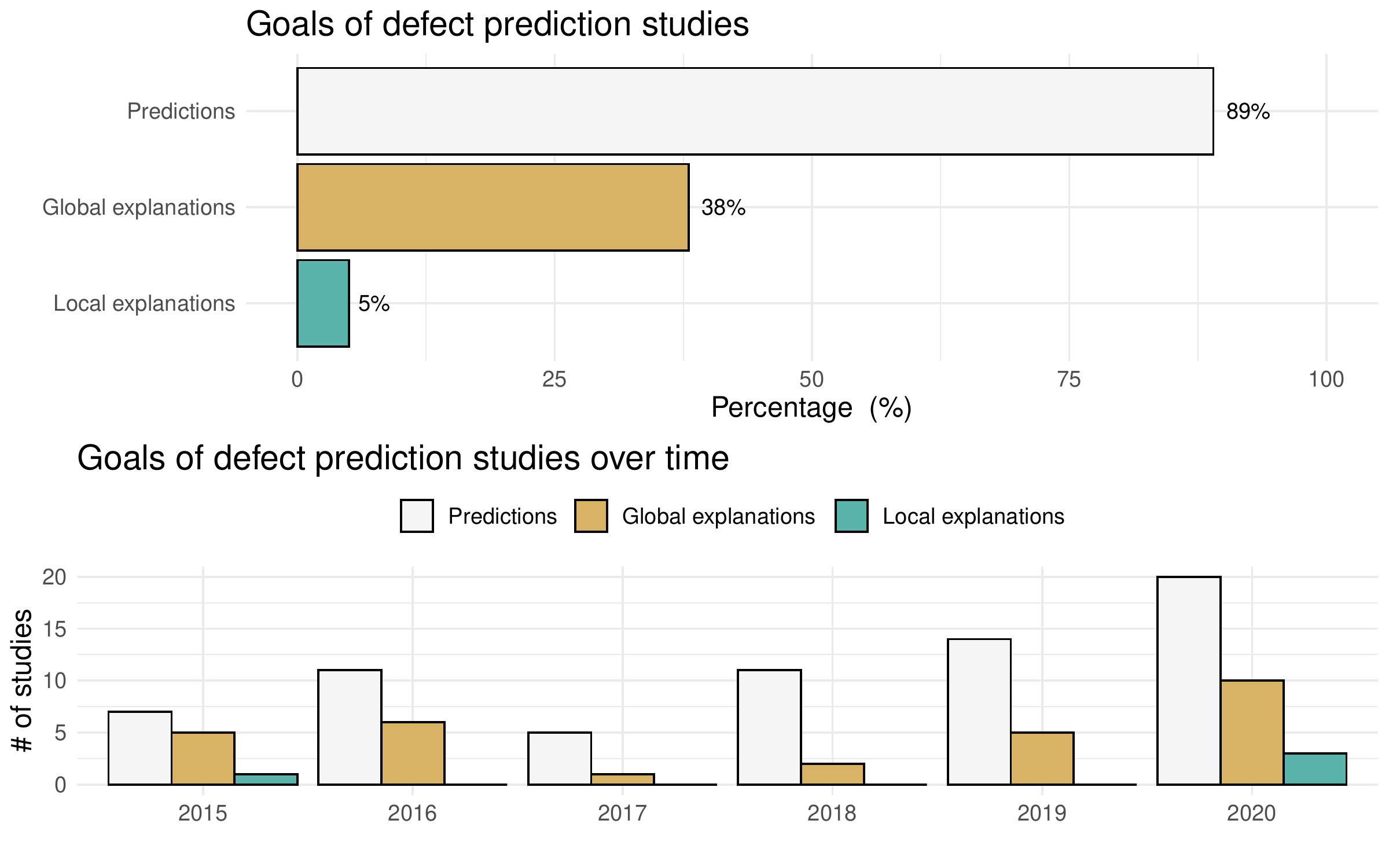}
  \caption{The results of our literature analysis indicate that research using explainable AI techniques to explain the predictions of defect prediction models remains largely unexplored (i.e., only 5\% of the 76 defect prediction studies).}
  \label{figure-survey}
\end{figure*}

Today software development processes depend on a variety of development tools (e.g., issue tracking systems, version control systems, code review, continuous integration, continuous deployment, and Q\&A website). 
Such tools generate large quantities of unstructured software artefacts at a high frequency (so-called Big Data) in many forms like issue reports, source code, test cases, code reviews, execution logs, app reviews, developer mailing lists, and discussion threads~\cite{hassan2008road}.

AI4SE (or Software Analytics) is a sub-field in software engineering that focuses on leveraging AI/ML and data \emph{analytics} techniques to uncover interesting and actionable knowledge from the unprecedented amount of \emph{software data}~\cite{menzies2013software, tantithamthavorn2018pitfalls}. 
Many software organisations (e.g., Microsoft, Facebook, and Google) currently use powerful AI/ML techniques to make data-driven engineering decisions and support software engineering tasks~\cite{tantithamthavorn2016automated,agrawal2018better, Bird2009, zimmermann2005mining, tantithamthavorn2018impact}.
For example, software defect prediction, effort estimation, task prioritization, task recommendation, expert recommendation, code generation, developers' productivity prediction, malware detection, and security vulnerability detection.

\subsection{On the Needs of Explainable AI for Software Engineering} 

While the adoption of software analytics systems enables software organisations to distill actionable insights and support decision-making, there are still many barriers to the successful adoption of such software analytics systems in software companies~\cite{dam2018explainable}. 
Below, we discuss three reasons why Explainable AI is needed in Software Engineering.

First, software practitioners do not understand the reason behinds the predictions from software analytics systems~\cite{dam2018explainable}. 
They often ask the following questions:
\begin{itemize}
    \item Why is this person best suited for this task?
    \item Why is this file predicted as defective?
    \item Why is this task required the highest development effort?
    \item Why should this task be done first?
    \item Why is a developer predicted to have low productivity?
    \item How can we improve the quality of software systems in the next iterations?
\end{itemize}
These concerns about a lack of explanations often to a lack of trust and hinder the adoption of such software analytics systems in practice.

\begin{figure*}[t]
   \centering
   \includegraphics[clip,width=\linewidth, trim={0 0 0 0}]{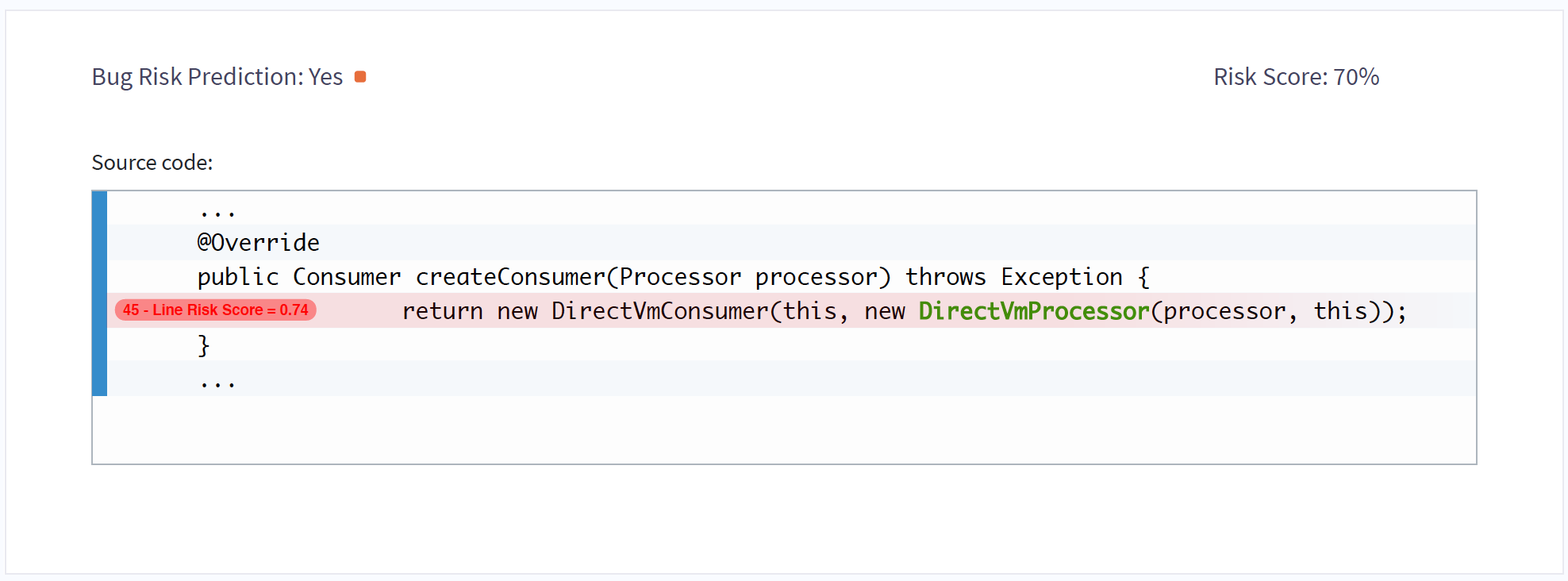}
   \caption{An example of the use of LIME to identify which lines of code are the most risky.}
   \label{figure-limeline}
\end{figure*}

Second, software practitioners are often affected by any decisions from these software analytics systems (e.g., would developers be laid-off since a defect prediction model found that a developer introduced software defects?).
Recently, Article 22 of the European Union’s General Data Protection Regulation (GDPR) states that the use of data in decision-making that affects an individual or group requires an explanation for any decision made by an algorithm. 
Unfortunately, current software analytics systems still do not uphold any privacy laws~\cite{jiarpakdee2020modelagnostic}.
Thus, the risks of unjustified decision-making of software analytics systems can be catastrophic, leading to potentially erroneous and costly business decisions~\cite{dam2018explainable}. 

Third, we find that as little as 5\% of the defect prediction studies focus on generating local explanations using explainable AI techniques (see Figure~\ref{figure-survey}).
We conduct a literature analysis to identify the popularity of defect prediction studies that adopt explainable AI techniques.
We first collect 76 primary defect prediction papers that were published in the top-tier Software Engineering venues (i.e., IEEE Transactions on Software Engineering, Empirical Software Engineering, and International Conference on Software Engineering) during 2015-2020 (as of October 1, 2020). 
Through a manual analysis, we found that: 89\% of the defect prediction studies only focus on the predictions, without considering generating explanations, while 38\% of the defect prediction studies focus on generating global explanations. 
Surprisingly, as little as 5\% of the defect prediction studies focus on generating local explanations using explainable AI techniques.
This insight suggests that explainable AI has been recently adopted in software engineering, but it is still under research.

\subsection{Towards Explainable AI for Software Engineering} 

To address these challenges, the overarching theme of Explainable AI for Software Engineering is concerned with the fundamental challenges of how can we leverage Explainable AI in the domain of software engineering to enhance the \textbf{practicality, explainability, and actionability} of software analytics.
Below, we demonstrate three successful case studies of using Explainable AI in Software Engineering to address the problem of software defect prediction models.

\section{EXPLAINABLE DEFECT PREDICTION MODELS: A CASE STUDY}

\subsection{Problem Motivation}
In today’s increasingly digitalized world, software defects are widespread and enormously expensive, but they are very hard to detect, predict, and prevent. 
Thus, a failure to eliminate software defects in safety-critical systems could result in serious injury to people, threats to life, death, and disasters. 

Traditionally, software quality assurance activities like software testing and code review are widely adopted to discover software defects in software systems. 
However, ultra-large-scale systems, such as, Google, can consist of more than two billion lines of code.
Thus, exhaustively reviewing and testing every single line of code is not feasible with limited time and resources.

\begin{figure*}[t]
   \centering
   \includegraphics[clip,width=\linewidth, trim={0 630 0 70}]{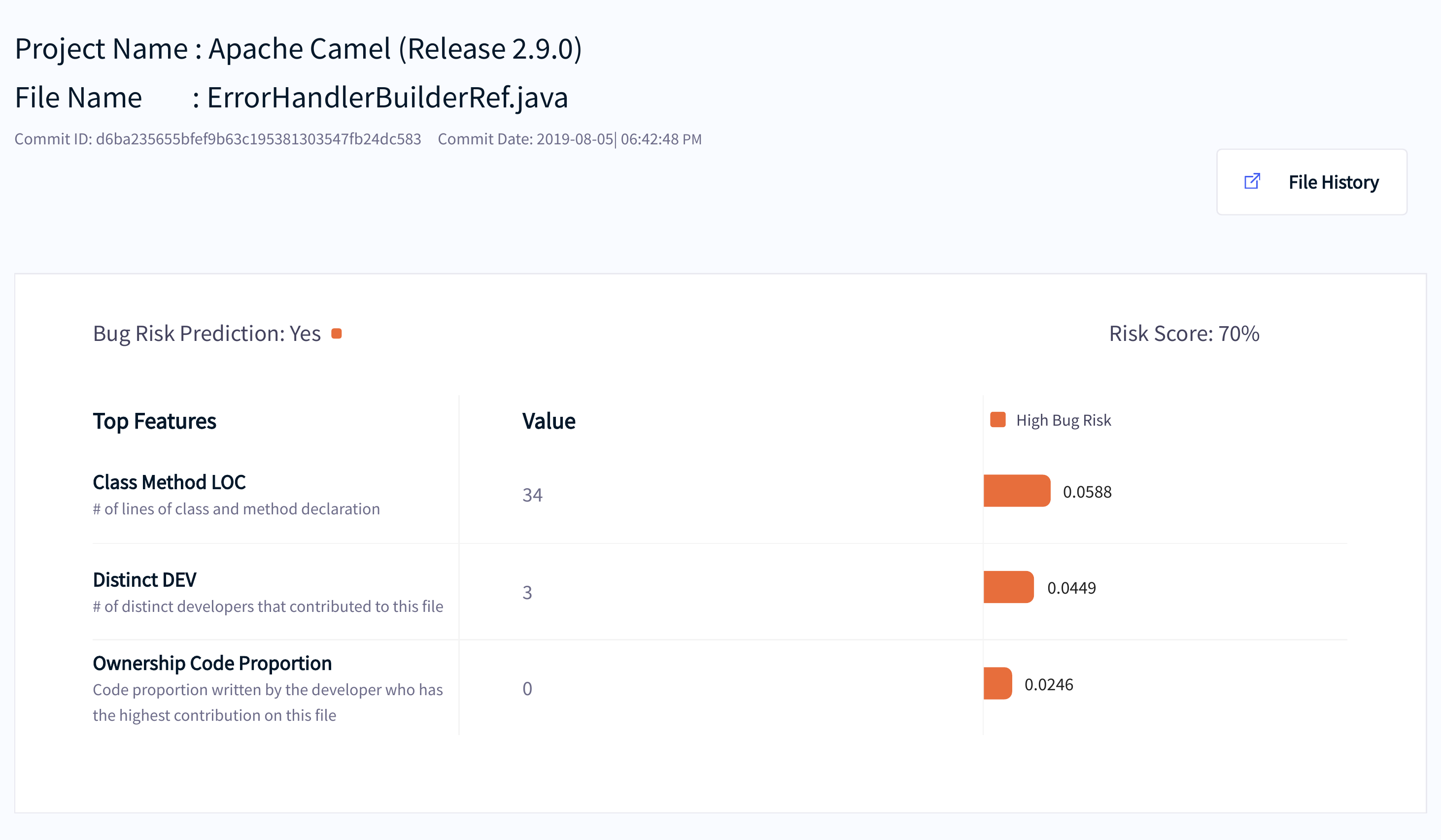}
   \caption{An example of an explanation generated by LIME to understand why a file is predicted as defective.}
   \label{figure-lime}
\end{figure*}

\subsection{Goals}
To this end, we developed an explainable defect prediction framework to achieve the following goals:

\begin{enumerate}
    \item Generating fine-grained predictions in order to help developers localize which lines of code are the most risky so developers can allocate limited software quality assurance activities in a cost-effective manner~\cite{wattanakriengkrai2020}.
    \item Generating explanations for each prediction to help developers understand why a file is predicted as defective~\cite{jiarpakdee2020modelagnostic}.
    \item Generating actionable guidance to help managers chart appropriate quality improvement plans.
\end{enumerate}

\subsection{Help developers localize which lines of code are the most risky}

\noindent \textbf{Motivation.} 
Traditionally, software defect predictions are proposed to predict which files are likely to be defective in the future. 
However, the proportion of critically-important defective lines are extremely low.
Our prior studies found that the ratio of defective lines in a file is as low as 1\%-3\%.
Thus, traditional file-level defect prediction models are still not practical to be used in practice.
However, there exists no features at the line level. 
Thus, line-level defect prediction remains an extremely challenging problem.

\noindent \textbf{Approach.} 
To address this problem, we proposed to use the LIME's model-agnostic technique~\cite{ribeiro2016model} to explain the predictions of file-level defect prediction models.
In particular, we first build file-level defect prediction models using textual features (i.e., bag of tokens that appear in a file) with a random forest classification technique.
For each prediction, we apply LIME to understand the prediction.
This approach allows us to identify which tokens and which lines contribute to the prediction of each file, which can be used to help developers localize which lines of code are the most risky.

\noindent \textbf{Results.} 
Figure~\ref{figure-limeline} shows an example of the use of LIME to identify which lines of code are the most risky.
The results show that this approach can correctly identify 61\% of actual defective lines in a file, suggesting our approach could potentially help developers reduce SQA effort that need to be spent by 52\% on clean lines, while accurately identifying 61\% of actual defective lines.

\begin{figure*}[t]
   \centering
   \includegraphics[clip,width=.95\linewidth, trim={0 340 0 70}]{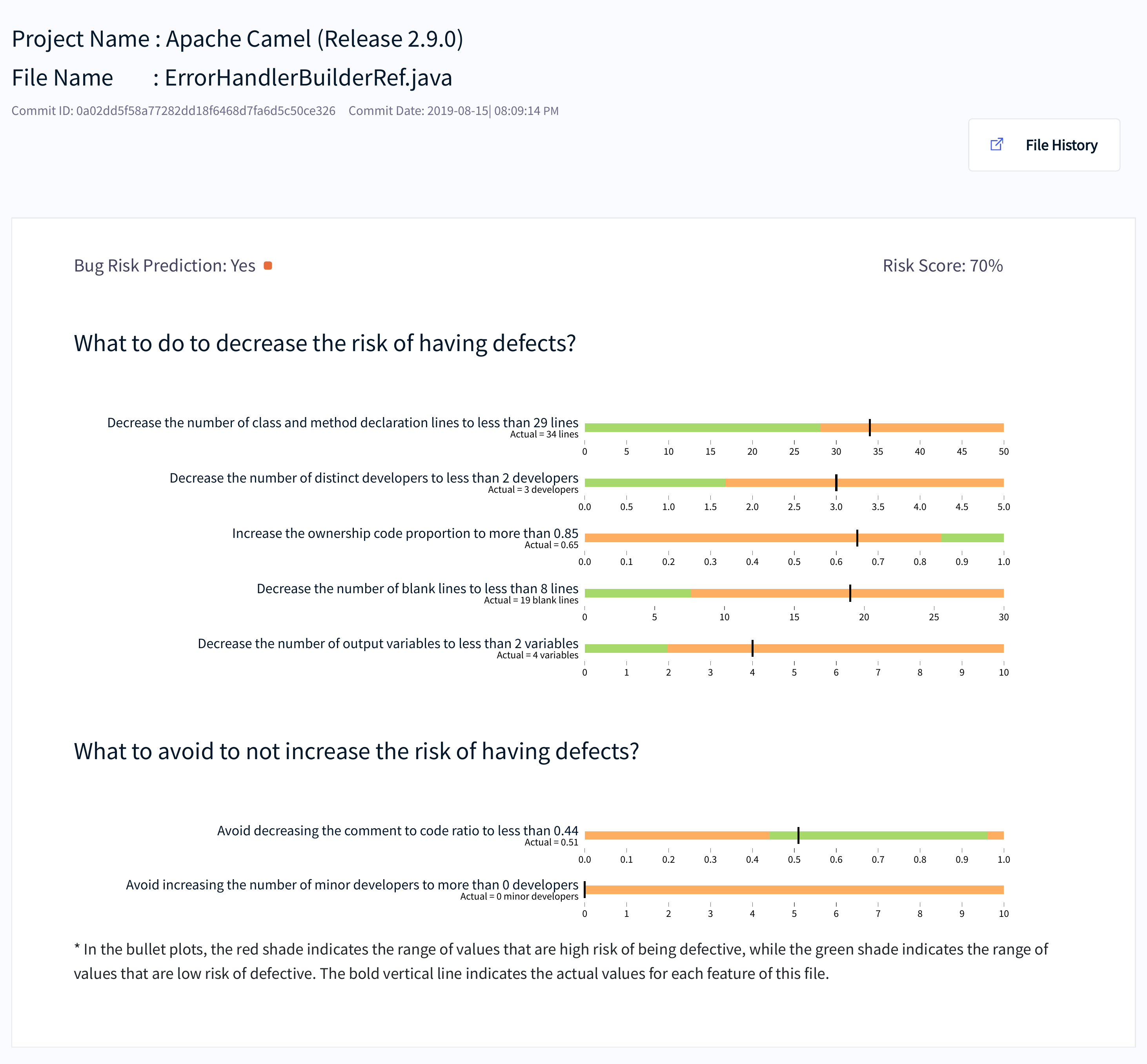}
   \caption{An example of actionable guidance to help managers developing quality improvement plans.}
   \label{fig:rules}
\end{figure*}

\subsection{Help developers understand why a file is predicted as defective}

\noindent \textbf{Motivation.} 
Traditionally, the predictions of defect models can help developers prioritize which files are the most risky.
However, developers do not understand why a file is predicted as defective, leading to a lack of trust in the predictions and hindering the adoption of defect prediction models in practice.
Thus, a lack of explainability of defect prediction models remains an extremely challenging problem.

\noindent \textbf{Approach.} 
To address this problem, we proposed to use a model-agnostic technique called LIME~\cite{ribeiro2016model} to explain the predictions of file-level defect prediction models.
In particular, we first build file-level defect prediction models that are trained using traditional software features (e.g., lines of code, code complexity, the number of developers who edited a file) with a random forest classification technique.
For each prediction, we apply LIME to understand the prediction.
This approach allows us to identify which features contribute to the prediction of each file.
This will help developers understand why a file is predicted as defective.

\noindent \textbf{Results.} 
Figure~\ref{figure-lime} presents an example of a visual explanation generated by LIME to understand why a file is predicted as defective.
According to this visual explanation, this file is predicted as defective with a risk score of 70\%.
The top-3 important factors that support this prediction are (1) the high number of class and method declaration lines, (2) the high number of distinct developers that contributed to the file, and (3) the low proportion of code ownership.
Thus, to mitigate the risk of having defects for this file, developers should consider decreasing the number of class and method declaration lines, reducing the number of distinct developers, and increasing the proportion of code ownership.

\subsection{Help managers develop software quality improvement plans}

\noindent \textbf{Motivation.} 
Traditionally, software defect predictions can only generate predictions and provide an understanding at the global level through the use of ANOVA analysis or Random Forest's feature importance~\cite{jiarpakdee2018impact}. 
However, such predictions and global insights are still not actionable---i.e., developers do not know what actions they should follow or should not follow to improve the quality of software systems.

Let's consider a scenario where a defect prediction model indicates that large file size is associated with defect-proneness.
While this insight can help managers develop quality improvement plans that in the next development iteration, developers should maintain the lower file size to mitigate the risk of having defects.
However, such insights do not provide a concrete suggestion of the optimal threshold of file size---i.e., what is the optimal file size that a file should be.
Thus, a lack of such actionable guidance and a lack of optimal  thresholds often leads to ineffective software quality improvement plans.

\noindent \textbf{Approach.} 
To generate actionable guidance, we propose to use a rule-based model-agnostic technique to generate a rule-based explanation for each  prediction of defect prediction models.
In particular, we first build file-level defect prediction models that are trained using traditional software features (e.g., lines of code, code complexity, the number of developers who edited a file) with a random forest classification technique.
For each prediction, we apply a rule-based model-agnostic technique called LoRMikA\cite{rajapaksha2020lormika} to generate two types of actionable guidance (i.e., what developers should do to mitigate the risk of having defects and what developers should not do to avoid increasing the risk of having defects).

\noindent \textbf{Results.} 
Figure~\ref{fig:rules} presents an example of actionable guidance to help managers developing quality improvement plans.
To decrease the risk of having defects, developers should consider (1) decrease the number of class and method declaration lines to less than 29 lines, (2) decrease the number of distinct developers to less than 2 developers, (3) increase the proportion of code ownership to more than 0.85, (4) decrease the number of blank lines to less than 8 lines, (5) decrease the number of output variables to less than 2 variables.

Nevertheless, to not increase the risk of having defects, developers should consider avoid decreasing the comment to code ratio and avoid increasing the number of minor or junior developers.





\section{CONCLUSION}

Based on the literature analysis on Explainable AI for SE and our successful case studies of Explainable AI for software defect prediction models, we draw the following conclusions:

\begin{enumerate}
    \item Explainable AI is very important in software engineering, but still under research, as shown in the literature analysis.
    \item Explainable AI techniques can be brought into software engineering to provide explanations of the predictions and actionable guidance to support software engineering tasks, as shown by three successful case studies of defect prediction models---using a model-agnostic technique called LIME to help developers localize which lines of code are the most risky, explain the predictions of defect models, and using a rule-based model agnostic technique to generate actionable guidance on what developers should do or should not do to prevent software defects.
\end{enumerate}

\section{ACKNOWLEDGMENT}

We thank our collaborators for their joint work including Supatsara Wattanakriengkrai, Patanamon Thongtanunam, Hideaki Hata, Kenichi Matsumoto, Hoa Khanh Dam, Dilini Rajapaksha, Christoph Bergmeir, and Wray Buntine.

C. Tantithamthavorn was partially supported by the Australian Research Council's Discovery Early Career Researcher Award (ARC DECRA) funding scheme (DE200100941).
J. Grundy was partially supported by the Australian Research Council's Laureate Fellowship funding scheme (FL190100035).

\bibliographystyle{IEEEtranS}
\bibliography{filteredref}

\end{document}